\documentclass[twocolumn,aps,superscriptaddress,showpacs]{revtex4}

\usepackage{amssymb}
\usepackage{amsmath}
\usepackage{graphicx}
\usepackage[normalem]{ulem}
\usepackage[dvips]{color}

\begin{document}

\title{Transition density and pressure in hot neutron stars}
\author{Jun Xu}
\affiliation{Cyclotron Institute, Texas A\&M University, College
Station, TX 77843-3366, USA}
\author{Lie-Wen Chen}
\affiliation{Department of Physics, Shanghai Jiao Tong University,
Shanghai 200240, China} \affiliation{Center of Theoretical Nuclear
Physics, National Laboratory of Heavy Ion Accelerator, Lanzhou
730000, China}
\author{Che Ming Ko}
\affiliation{Cyclotron Institute and Department of Physics and
Astronomy, Texas A\&M University, College Station, TX 77843-3366,
USA}
\author{Bao-An Li}
\affiliation{Department of Physics and Astronomy, Texas A\&M
University-Commerce, Commerce, TX 75429-3011, USA}

\begin{abstract}

Using the momentum-dependent MDI effective interaction for nucleons,
we have studied the transition density and pressure at the boundary
between the inner crust and liquid core of hot neutron stars.  We
find that their values are larger in neutrino-trapped neutron stars
than in neutrino-free neutron stars. Furthermore, both are found to
decrease with increasing temperature of a neutron star as well as
increasing slope parameter of the nuclear symmetry energy, except
that the transition pressure in neutrino-trapped neutron stars for
the case of small symmetry energy slope parameter first increases
and then decreases with increasing temperature.  We have also
studied the effect of the nuclear symmetry energy on the critical
temperature above which the inner crust in a hot neutron star
disappears and found that with increasing value of the symmetry
energy slope parameter, the critical temperature decreases slightly
in neutrino-trapped neutron stars but first decreases and then
increases in neutrino-free neutron stars.

\end{abstract}

\pacs{26.60.-c, 
      21.30.Fe, 
      21.65.-f, 
      97.60.Jd  
      }

\maketitle

\section{Introduction}\label{introduction}

Studying the properties of neutron stars allows us to test our
knowledge on the properties of nuclear matter under extreme
conditions. Theoretical studies have shown that a neutron star is
expected to have a liquid core surrounded by an inner
crust~\cite{Cha08}, which extends outward to the neutron drip-out
region. While the neutron drip-out density $\rho _{\rm out}$ has
been relatively well determined~\cite{Rus06}, the transition density
$\rho _{t}$ at the inner edge of the crust is still quite uncertain
because of our limited knowledge on the nuclear equation of state
(EOS), especially the density dependence of the symmetry energy
($E_{\rm sym}(\rho)$) of neutron-rich nuclear
matter~\cite{Lat00,Lat07}. Recently, significant progress has been
made in constraining the EOS of neutron-rich nuclear matter using
terrestrial laboratory experiments (See Ref.~\cite{LCK08} for a
recent review).  In particular, from analyses of experimental data
on neutron skin thickness, isobaric analogue states, Pygmy dipole
resonances, and giant dipole resonances in nuclei as well as on
isospin diffusion, isoscaling, and neutron-proton to triton-$^3$He
ratio in intermediate-energy nuclear reactions, significant
constraints on $E_{\rm sym}(\rho)$ have been obtained for the same
sub-saturation density region as expected in the inner edge of
neutron star crusts. The extracted slope parameter $L=3\rho_0
(\partial E_{\rm sym}(\rho)/\partial\rho)_{\rho=\rho_0}$ of the
nuclear symmetry energy from these studies has values in the range
$30$ MeV  $< L < 80$ MeV~\cite{She10}. With the MDI interaction
together with the value $L=86\pm25$ MeV constrained from an analysis
of the isospin-diffusion data~\cite{Tsa04,Che05a,LiBA05,Tsa09} in
heavy-ion collisions using the isospin-dependent
Boltzmann-Uehling-Uhlenbeck (IBUU) transport model with the
momentum-dependent MDI interaction~\cite{Das03}, the density and
pressure at the inner edge of the crust of cold neutron stars were
studied by considering the boundary of the instability region or the
spinodal boundary between the liquid core and inner crust of a cold
neutron star in both the thermodynamical approach~\cite{Kub07,Lat07}
and the dynamical approach~\cite{BPS71,BBP71,Pet95a,Pet95b,Oya07}.
This leads to the constraints $0.040$ fm$^{-3}<\rho_t<0.065$
fm$^{-3}$ and $0.01$ MeV/fm$^{3} < P_t < 0.26$ MeV/fm$^{3}$,
respectively, for the transition density and pressure. Together with
the crustal fraction of the total moment inertia of the Vela pulsar
extracted from its glitches~\cite{Lin99}, a tighter constraint on
the mass-radius relation of cold neutron stars was
obtained~\cite{XCLM09}.

Because of the initial high temperature and appreciable proton
fraction in a newly-formed neutron star immediately after
gravitational collapse of a massive star~\cite{Bur86,Bet90,Bur03},
neutrinos are abundantly produced from the Urca process in its inner
core. Although high energy neutrinos can be trapped at densities as
low as 10$^{12}$ g/cm$^3$~\cite{BBLA79}, the stars cools by neutrino
emissions. As neutrinos are emitted from this so-called
proto-neutron star, which has an initial temperature of $\sim
10^{11}K$ (about $10$ MeV)~\cite{Bur88,Hor04}, its temperature drops
to $\sim 10^{10} K$ (about $1$ MeV) and even lower.  Afterwards, the
neutron star becomes transparent to neutrinos as their mean free
path increases with decreasing energy, and the cooling of the
neutron star continues to be dominated by neutrino emission for a
long time. It is thus of interest to study the transition density
and pressure in newly-born hot neutron stars, as this would help to
understand the cooling mechanism and structural evolution of neutron
stars. In this paper, we extend the study of Ref.~\cite{XCLM09} to
finite temperature and study the dependence of the transition
density and pressure of hot neutron stars on the nuclear symmetry
energy, particularly its slope at nuclear saturation density.

This paper is organized as follows. We first review in
Sec.~\ref{model} the momentum-dependent MDI interaction for
nucleons, in Sec.~\ref{nsmatter} the properties of hot neutron star
matter, and in Sec.~\ref{approaches} the dynamical approach for
locating the inner edge of the crust of a hot neutron star. We then
show in Sec.~\ref{results} the results and conclude with a summary
in Sec.~\ref{summary}.

\section{The MDI interaction}\label{model}

The MDI interaction is an effective nuclear interaction with its
density and momentum dependence constrained from the
phenomenological finite-range Gogny interaction~\cite{Das03}. In the
mean-field approximation, the potential energy density of a nuclear
matter of density $\rho$ and isospin asymmetry
$\delta=(\rho_n-\rho_p)/\rho$ , with $\rho_n$ and $\rho_p$ being,
respectively, the neutron and proton densities, can be expressed
as~\cite{Das03,Che05a}
\begin{eqnarray}
V(\rho,\delta ) &=&\frac{A_{u}(x)\rho _{n}\rho
_{p}}{\rho
_{0}} +\frac{A_{l}(x)}{2\rho _{0}}(\rho _{n}^{2}+\rho_{p}^{2})\notag\\
&+&\frac{B}{\sigma +1}\frac{\rho ^{\sigma +1}}{\rho
_{0}^{\sigma }}(1-x\delta ^{2})\notag\\
&+&\frac{1}{\rho _{0}}\sum_{\tau ,\tau^{\prime}}C_{\tau ,\tau ^{\prime }}
\int \int d^{3}pd^{3}p^{\prime }\frac{f_{\tau }(\vec{r},\vec{p}%
)f_{\tau ^{\prime }}(\vec{r}^{\prime },\vec{p}^{\prime
})}{1+(\vec{p}-\vec{p}^{\prime })^{2}/\Lambda ^{2}}.\notag\\
\label{MDIVB}
\end{eqnarray}%
In the above equation, $\tau(\tau^\prime)$ is the nucleon isospin
taken to be $1/2$ for neutron and $-1/2$ for proton;
$f_{\tau}(\vec{r},\vec{p})=(2/h^{3})\{\exp [(p^2/2m+U_{\tau }-\mu
_{\tau })/T]+1\}^{-1}$ is the nucleon phase-space distribution
function in a thermally equilibrated nuclear matter with $m=939$ MeV
being the nucleon mass, $\mu_\tau$ being the chemical potential of
nucleon of isospin $\tau$, $T$ being the temperature and $U_\tau$
being the nucleon mean-field potential to be introduced below; and
$\rho_0=0.16$ fm$^{-3}$ is the saturation density of normal nuclear
matter. Values of the parameters $A_u(x)$, $A_l(x)$, $B$, $\sigma$,
$\Lambda$, $C_l=C_{\tau,\tau}$ and $C_u=C_{\tau,-\tau}$ can be found
in Refs.~\cite{Das03,Che05a}. For symmetric nuclear matter, this
interaction gives a binding energy of $-16$ MeV per nucleon and an
incompressibility $K_0$ of $212$ MeV at saturation density.

Taking the derivative of Eq.~(\ref{MDIVB}) with respect to the
proton or neutron density leads to the following single-particle
potential for a nucleon of isospin $\tau $:
\begin{eqnarray}
U(\rho,\delta ,\vec{p},\tau ) &=&A_{u}(x)\frac{\rho _{-\tau }}{\rho _{0}}%
+A_{l}(x)\frac{\rho _{\tau }}{\rho _{0}}\notag\\
&+&B\left(\frac{\rho }{\rho _{0}}\right)^{\sigma }(1-x\delta ^{2})\notag\\
&-&8\tau x\frac{B}{\sigma +1}\frac{\rho ^{\sigma -1}}{\rho _{0}^{\sigma }}
\delta \rho_{-\tau }\notag\\
&+&\frac{2C_{\tau ,\tau }}{\rho _{0}} \int d^{3}p^{\prime }\frac{f_{\tau }(%
\vec{r},\vec{p}^{\prime })}{1+(\vec{p}-\vec{p}^{\prime
})^{2}/\Lambda ^{2}}\notag\\
&+&\frac{2C_{\tau ,-\tau }}{\rho _{0}} \int d^{3}p^{\prime }\frac{f_{-\tau }(%
\vec{r},\vec{p}^{\prime })}{1+(\vec{p}-\vec{p}^{\prime
})^{2}/\Lambda ^{2}}, \label{MDIU}
\end{eqnarray}%
which is seen to depend on the momentum $\vec{p}$ of the nucleon.
Although the properties of cold nuclear matter can be essentially
determined analytically using Eqs.~(\ref{MDIVB}) and (\ref{MDIU}),
these equations needs to be solved numerically and self-consistently
by the iteration method~\cite{Xu07} to obtain the thermodynamical
quantities of hot nuclear matter.

In cold nuclear matter, the symmetry energy from the MDI interaction
is given  by
\begin{eqnarray}\label{esymmdi}
E_{\rm sym}(\rho) &=& \frac{1}{2} \left(\frac{\partial^2 E}
{\partial \delta^2}\right)_{\delta=0} \notag\\
&=& \frac{8 \pi}{9 m h^3 \rho} p^5_f + \frac{\rho}{4 \rho_0}
(-24.59+4Bx/(\sigma +1)) \notag\\
&-& \frac{B x}{\sigma + 1} \left(\frac{\rho}{\rho_0}\right)^\sigma +
\frac{C_l}{9 \rho_0 \rho} \left(\frac{4 \pi}{h^3}\right)^2 \Lambda^2 \notag\\
&\times&  \left[4 p^4_f - \Lambda^2 p^2_f \ln \left(\frac{4 p^2_f
+ \Lambda^2}{\Lambda^2}\right)\right] \notag\\
&+& \frac{C_u}{9 \rho_0 \rho} \left(\frac{4 \pi}{h^3}\right)^2
\Lambda^2 \notag\\
&\times& \left[4 p^4_f - p^2_f (4 p^2_f + \Lambda^2) \ln
\left(\frac{4
p^2_f + \Lambda^2}{\Lambda^2}\right)\right],\notag\\
\end{eqnarray}
where $p_f=\hbar(3\pi^2\rho/2)^{1/3}$ is the nucleon Fermi momentum
in symmetric nuclear matter. The first term in Eq.~(\ref{esymmdi})
is the contribution from the kinetic part while other terms are from
the potential part. The symmetry energy is fixed to be $30.5$ MeV at
normal nuclear density, and the parameter $x$ is used to model the
density dependence of the symmetry energy away from the saturation
density without changing the properties of symmetric nuclear matter.
The resulting slope parameter of the symmetry energy has values of
about $15$ MeV, $60$ MeV, and $106$ MeV for $x=1$, $0$, and $-1$, respectively.

\begin{figure}[h]
\centerline{\includegraphics[scale=0.9]{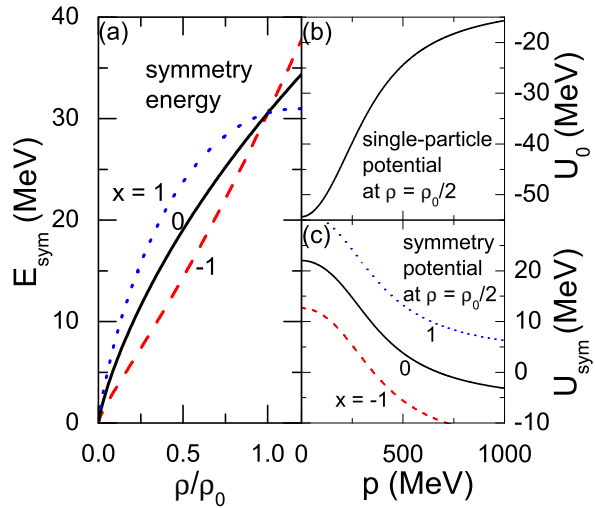}} \caption{(Color
online) (a) Density dependence of symmetry energies, (b)
single-particle potentials in symmetric nuclear matter and (c)
symmetry potentials in  cold nuclear matter at $\rho_0/2$ for the
MDI interaction.} \label{EsymU}
\end{figure}

The density dependence of the symmetry energy from the MDI
interaction is shown in panel~(a) of Fig.~\ref{EsymU} for $x=1$,
$x=0$ and $x=-1$. With $x=1$ ($x=-1$) the symmetry energy is larger
(smaller) at subsaturation densities but smaller (larger) at
suprasaturation densities, and it is called a 'soft' ('stiff')
symmetry energy. For $x=0$, the value of the symmetry energy lies
between those of $x=1$ and $x=-1$. In panels~(b) and (c), the
momentum dependence of the nucleon single-particle potential $U_0$
in symmetric nuclear matter at half saturation density is displayed.
Also shown in the figure is the symmetry potential, defined as
$U_{\rm sym}=(U_n-U_p)/2\delta$ and calculated at $\delta=0.2$ and
the same density. It is seen that $U_0$ has negative values at low
momenta but increases with increasing momentum and saturates at high
momenta. The symmetry potential $U_{\rm sym}$, which measures the
difference between the neutron and proton single-particle potentials
in an asymmetric nuclear matter, decreases with increasing momentum
and is larger for $x=1$ than for $x=0$ and $x=-1$, which is
consistent with the behavior of the symmetry energy at subsaturation
densities.

\section{Hot neutron star matter}
\label{nsmatter}

For a newly-born neutron star, its cooling is dominated by the
emission of neutrinos, which are produced through the Urca process,
leading to the following charge neutrality and $\beta$-stable
conditions:
\begin{eqnarray}
\rho_p &=& \rho_e,\\
\mu_p + \mu_e &=& \mu_n + \mu_{\nu_e}.
\end{eqnarray}
The production of muons is negligible for the density and
temperature present in the hot neutron star crust. Neutrino trapping
in the early stage of a supernova has been extensively studied in
the literature~\cite{BBLA79}, and it was found that the fraction of
leptons
\begin{equation}
Y_l = \frac{\rho_e+\rho_{\nu_e}}{\rho}
\end{equation}
is about $0.35 \sim 0.4$ at the onset of trapping~\cite{Bur88,Bet90}
and decreases as neutrinos leave the star. In the present study, we
consider the two extreme cases of neutrino-trapped and neutrino-free
hot neutron star matters. In the former case, we choose $Y_l$ to be
$0.4$ as an example and consider the typical temperature of $5$ or
$10$ MeV. In the case of neutrino-free hot neutron star matter, we
set the neutrino chemical potential to be $0$ and the temperature to
be $0$ or $1$ MeV, corresponding to the later stage of a neutron
star's evolution. The abundance of each species in both cases can be
calculated from above equations together with the baryon number
conservation condition $\rho=\rho_n+\rho_p$.

\begin{figure}[h]
\centerline{\includegraphics[scale=0.9]{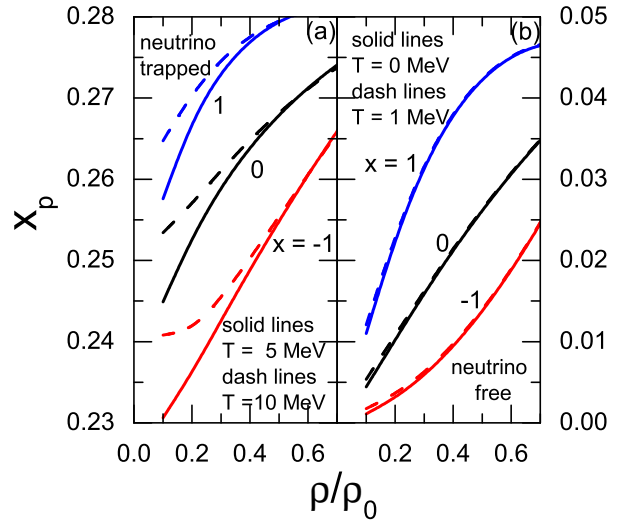}} \caption{(Color
online) The proton fraction $x_p$ as a function of baryon density $\rho$
from the MDI interaction for both neutrino-trapped matter (left panel) and
neutrino-free matter (right panel) with different $x$ parameters and
at different temperatures. Note that different scales for
$x_p$ are used for the neutrino-free
and the neutrino-trapped matter.} \label{rhodelta}
\end{figure}

Assuming that both electrons and neutrinos are massless, we
determine the proton fraction $x_p=(1-\delta)/2$ in both the
neutrino-trapped matter and the neutrino-free matter as a function
of baryon density $\rho$, and the results are shown in
Fig.~\ref{rhodelta}. It is seen that the neutrino-free matter is
much more neutron-rich than the neutrino-trapped matter. The
critical proton fraction $11 \sim 15$ \% for the direct Urca
process~\cite{Bog79,Lat91} is smaller than the proton fraction in
the neutrino-trapped matter but larger than the proton fraction in
the neutrino-free matter. The proton fraction increases with
increasing density in all cases and slightly increases with
increasing temperature for a fixed density, especially at low
densities. These can be understood from the increase of the symmetry
free energy with increasing temperature and density~\cite{Xu07},
which makes the neutron star matter more symmetric at higher
temperatures and densities. Furthermore, the stiff symmetry energy
($x=-1$) makes the system more neutron-rich at subsaturation
densities as expected.

The total pressure $P$ in a hot neutron star matter can be written as
\begin{equation}
P = P_b + P_l,
\end{equation}
where the baryon contribution $P_b(\rho ,T,\delta )$ is calculated
from the thermodynamic relation
\begin{eqnarray}
P_b(\rho ,T,\delta ) &=&\left[ T{\sum_{\tau }}s_{\tau }(\rho
,T,\delta)-V(\rho ,T,\delta )\right. \notag\\
&&\left.- V_{\rm kin}(\rho,T,\delta)\right]+\sum_{\tau }\mu _{\tau }\rho _{\tau }.  \label{Pb}
\end{eqnarray}
In the above equation, $V(\rho ,T,\delta )$ and $V_{\rm
kin}(\rho,T,\delta)$ are, respectively, the potential and kinetic
contributions to the total energy density with the latter given by
\begin{equation}
V_{\rm kin}(\rho,T,\delta) = {\sum_{\tau }}%
\int d^{3}p\frac{p^{2}}{2m}f_{\tau }(\vec{r},\vec{p}),
\end{equation}
and $s_{\tau }(\rho ,T,\delta )$ is the entropy density, which is given by
\begin{equation}
s_{\tau }(\rho ,T,\delta )=-\frac{8\pi }{{ }h^{3}}\int_{0}^{\infty
}p^{2}[n_{\tau }\ln n_{\tau }+(1-n_{\tau })\ln (1-n_{\tau })]dp,
\label{S}
\end{equation}%
with the particle occupation number%
\begin{equation}
n_{\tau }=\frac{1}{\exp [(p^{2}/2m+U_{\tau }-\mu _{\tau })/T]+1}.
\end{equation}%
Above formula can also be used for the calculation of $P_l$ by using
$l$ ($l=e, \nu_e$) instead of $\tau$, and leptons are treated as
non-interacting ultra-relativistic particles.

\section{Locating the inner edge of the neutron star crust}
\label{approaches}

We briefly review in this section the dynamical approach that will
be used to locate the inner edge of the neutron star
crust~\cite{Cha08}, and discuss its application to the case of
finite temperature. We neglect muons in the neutron star crust as
their number is small compared to that of electrons at low densities
and temperatures.

In the dynamical approach, the stability condition for a homogeneous
neutron star matter against small periodic density perturbations can be
well approximated by \cite%
{BPS71,BBP71,Pet95a,Pet95b,Oya07}
\begin{equation}
V_{\rm dyn}(k)=V_{0}+\beta k^{2}+\frac{4\pi
e^{2}}{k^{2}+k_{TF}^{2}}>0, \label{Vdyn}
\end{equation}%
where $k$ is the wavevector of the spatially periodic density
perturbations and
\begin{eqnarray}
V_{0} &=&\frac{\partial \mu _{p}}{\partial \rho
_{p}}-\frac{(\partial \mu
_{n}/\partial \rho _{p})^{2}}{\partial \mu _{n}/\partial \rho _{n}},\text{ }%
k_{TF}^{2}=\frac{4\pi e^{2}}{\partial \mu_e /\partial \rho _{e}},  \notag \\
\beta  &=&D_{pp}+2D_{np}\zeta +D_{nn}\zeta ^{2},~~\zeta
=-\frac{\partial \mu _{n}/\partial \rho _{p}}{\partial \mu
_{n}/\partial \rho _{n}}.\notag\\
\end{eqnarray}%
The three terms in Eq.~(\ref{Vdyn}) represent, respectively,
contributions from the bulk nuclear matter, the density-gradient
terms, and the Coulomb interaction. The empirical values for the
coefficients of density-gradient terms are
$D_{pp}=D_{nn}=D_{np}=132$ MeV$\cdot $fm$^{5}$~\cite{Oya07,XCLM09}.
At $k_{\min }=[(4\pi e^2/\beta )^{1/2}-k_{TF}^{2}]^{1/2}$, $V_{\rm
dyn}(k)$ has the minimal value of $V_{\rm dyn}(k_{\min
})=V_{0}+2(4\pi e^{2}\beta )^{1/2}-\beta
k_{TF}^{2}$~\cite{BPS71,BBP71,Pet95a,Pet95b,Oya07}, and $\rho _{t}$
is then determined from $V_{\rm dyn}(k_{\min })=0$. We note that the
first term in Eq.~(\ref{Vdyn}) gives the dominant contribution in
the determination of the transition density, and including other
terms lowers the transition density. Also, although at low
temperatures the electron density $\rho_e$ can be written as an
expansion of the electron chemical potential
\begin{equation}\label{rhoelT}
\rho_e \approx \frac{8\pi}{3h^3} \mu_e^3 [1+\pi^2(T/\mu_e)^2],
\end{equation}
which gives an analytical expression for $\partial \mu_e/\partial
\rho_e$, at high temperatures numerical calculations are needed.

The dynamical approach reduces to the so-called thermodynamical
approach~\cite{Kub07,Lat07} in the long wave length limit when the
density gradient terms and the Coulomb interaction are
neglected~\cite{Mar03,XCLM09}, which leads to the stability
condition
\begin{equation}\label{ther}
V_{\rm ther}=\frac{\partial \mu _{p}}{\partial \rho
_{p}}-\frac{(\partial \mu _{n}/\partial \rho _{p})^{2}}{\partial \mu
_{n}/\partial \rho _{n}}>0.
\end{equation}

\begin{figure}[h]
\centerline{\includegraphics[scale=0.9]{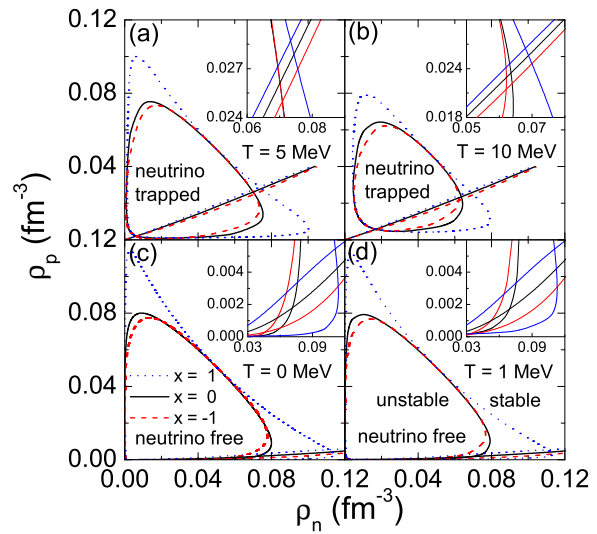}} \caption{(Color
online) The instability region and the relative neutron-proton
abundance of hot neutron star matter for different temperatures and
nuclear symmetry energy parameters. Regions where they cross each
other are shown in the insets in enlarged scales.} \label{rhonrhop}
\end{figure}

To illustrate the relation between the transition density and the
area of the spinodal region, we show in Fig.~\ref{rhonrhop} the
instability region of nuclear matter with the boundary determined by
$V_{\rm ther}=0$ and the relative neutron-proton abundance of a hot
neutron star in the $(\rho_n,\rho_p)$ plane. The cross point, also
shown with enlarged scales in the insets, is the transition density
from the thermodynamical approach. It is seen that although the
neutron star matter becomes less neutron-rich with increasing
temperature, the area of the instability region shrinks more quickly
with increasing temperature. Furthermore, the stiffness of the
symmetry energy also affects the shape and area of the spinodal
region. As temperature increases, the spinodal boundaries from
different values of $x$ cross with the relative neutron-proton
abundance curves at decreasingly small proton densities. For a more
detailed discussion on the symmetry energy and temperature effects
on the spinodal region, we refer readers to
Refs.~\cite{LiBA01,LiBA97b,Bar98,Duc07,Duc08,Ava04}. The following
analysis of the transition density and pressure in hot neutron stars
is, however, carried out by using the more realistic dynamical
approach.

\section{Results and discussions}
\label{results}

In this section, we show the temperature dependence of the
transition density and pressure in newly-born hot neutron stars by
using the MDI interaction with different values for the symmetry
energy parameter $x$ or the slope parameter $L$ of the symmetry
energy.

\begin{figure}[h]
\centerline{\includegraphics[scale=0.9]{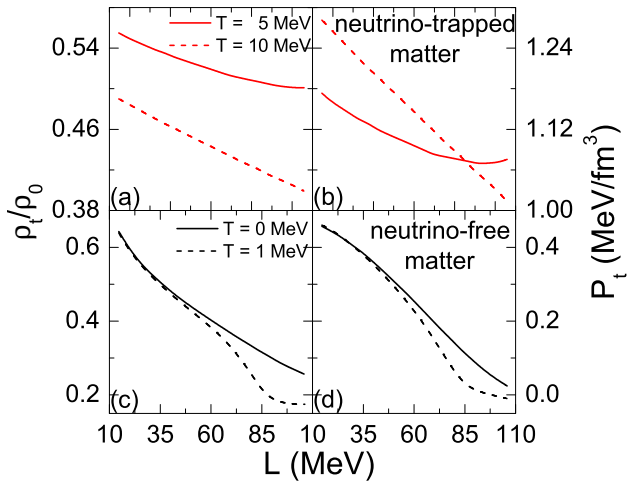}} \caption{(Color
online) Transition densities $\rho_t$ ((a) and (c)) and pressure $P_t$
((b) and (d)) as functions of the slope parameter $L$ of the
symmetry energy at different temperatures for both the
neutrino-trapped matter ((a) and (b)) and the neutrino-free matter
((c) and (d)).} \label{rhotL}
\end{figure}

The dependence of the transition density and pressure in hot neutron
stars on the slope parameter $L$ of the symmetry energy at different
temperatures is shown in Fig.~\ref{rhotL}. It is seen that the
transition density $\rho_t$ generally decreases with increasing
value of the slope parameter $L$ of the symmetry energy. As the
transition density can be viewed approximately as the beginning of a
first-order liquid-gas phase transition, a stiffer symmetry energy,
which corresponds to a softer equation of state at subsaturation
densities, leads thus to a smaller phase transition density and
therefore a lower core-crust transition density. Furthermore, the
transition density decreases with increasing temperature, and for
the neutrino-free matter this is more pronounced for larger values
of $L$. The temperature effect can again be understood by the
decreasing phase transition density with increasing temperature. For
the neutrino-trapped matter, the $L$-dependence of the transition
density is relatively weak. The weak $L$-dependence is mainly due to
the fact that the isospin asymmetry is not so large as shown in
Fig.~\ref{rhodelta}. We note that a similar temperature dependence
of the transition density has been obtained in studies based on
Skyrme interactions and relativistic mean field
models~\cite{Duc08,Par09,Ava09}. For the $L$-dependence of the
transition density, results from these models are, however, not so
clear as different values for the incompressibility $K_0$ of the
symmetric matter at saturation density and $E_{sym}(\rho_0)$, which
also affect the value of the transition density, have been used.

For the transition pressure $P_t$, its value in the neutrino-trapped
matter decreases only very slightly with increasing $L$ as a result
of the weak $L$-dependence of $\rho_t$. Its temperature dependence
shows, however, a complicated behavior of slightly higher and
smaller values at higher temperatures for smaller and larger values
of $L$, respectively. This is due to the fact that although the
transition density $\rho_t$ decreases with increasing temperature,
the contribution from leptons increases with increasing temperature.
For the neutrino-free matter, $P_t$ is seen to decrease rapidly with
increasing $L$. As to its temperature dependence, $P_t$ in the
neutrino-free matter decreases with increasing temperature at larger
values of $L$ but shows a weaker temperature dependence for smaller
values of $L$. Also, $P_t$ is larger for the neutrino-trapped matter
than for the neutrino-free matter. Interestingly, $P_t$ becomes very
small and even negative at $T=1$ with larger $L$ for the
neutrino-free matter. This is due to the smaller contributions from
the leptons and the asymmetric part of the nuclear interactions to
the total pressure. Since the pressure at the inner edge of neutron
star crust cannot be negative, our finding thus indicates that
either the neutrino-free matter in hot neutron stars cannot reach a
temperature above $T=1$ MeV or the symmetry energy cannot have a
slope parameter larger than $L\sim 100$ MeV.

\begin{figure}[h]
\centerline{\includegraphics[scale=0.9]{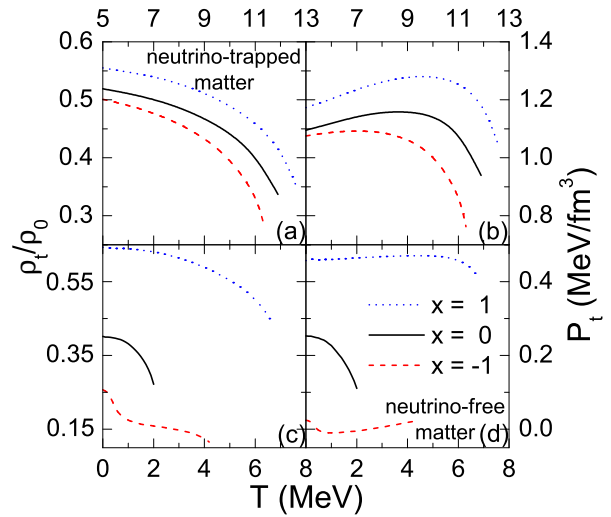}} \caption{(Color
online) Transition densities $\rho_t$ ((a) and (c)) and pressure $P_t$
((b) and (d)) as functions of temperature $T$ with $x=0$ and
$x=-1$ for both the neutrino-trapped matter ((a) and (b)) and
the neutrino-free matter ((c) and (d)). Note that different
scales for temperatures are used for the neutrino-trapped matter and
the neutrino-free matter.} \label{rhotT}
\end{figure}

The temperature effect on the transition density and pressure for a
fixed symmetry energy parameter is demonstrated in Fig.~\ref{rhotT}
for the symmetry energy parameters $x=1$, $x=0$ and $x=-1$. For the
neutrino-trapped matter, the temperature effect is similar for all
three $x$ values, and the transition density $\rho_t$ decreases
almost linearly with increasing temperature at lower $T$ and
decreases quickly at higher $T$. For the neutrino-free matter,
although the transition density $\rho_t$ decreases smoothly with
increasing temperature for $x=1$, it decreases slowly (quickly) at
lower temperatures with $x=0$ ($x=-1$) while quickly (slowly) at
higher temperatures. The temperature effects on the transition
density reflect those on the spinodal region and the abundance of
particle species in the hot neutron star matter. For the
neutrino-trapped matter, the transition pressure $P_t$ is seen to be
insensitive to the temperature for $x=-1$ but increases slightly
with increasing temperature for $x=0$ and $x=1$ at lower $T$, and it
decreases with increasing temperature at higher $T$ for all values
of $x$. For the neutrino-free matter, it is insensitive to
temperature for $x=1$ but decreases with increasing temperature for
$x=0$, while for $x=-1$ it drops to a negative value and becomes
positive again as the temperature increases. Our results again show
that the behavior of $P_t$ is dominated by that of $\rho_t$ and the
contribution from the leptons, with the former decreasing and the
latter increasing with increasing temperature.

\begin{figure}[h]
\centerline{\includegraphics[scale=0.9]{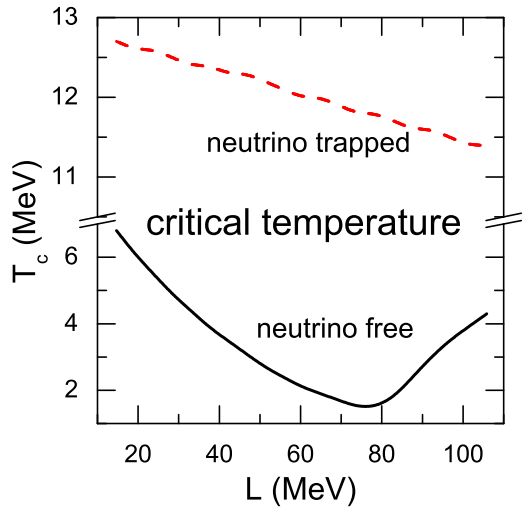}} \caption{(Color
online) Critical temperature $T_c$ as a function of the slope
parameters $L$ of the symmetry energy for both the neutrino-trapped
matter and the neutrino-free matter. } \label{rhotTc}
\end{figure}

We have seen in Fig.~\ref{rhonrhop} that as the temperature of the
neutron star matter increases, there will be eventually no cross
point between the curve of neutron-proton relative abundance and the
boundary of the spinodal region, leading to the disappearance of the
transition density in hot neutron stars. In such case, the inner
crust (nuclear 'pasta' phase) disappears and the liquid core expends
directly to the outer crust. To determine the critical temperature
$T_c$ at which the transition density $\rho_t$ disappears is thus
useful for understanding the structural evolution of newly-born hot
neutron stars. Figure~\ref{rhotTc} displays the $L$-dependence of
the critical temperature for both the neutrino-trapped matter and
the neutrino-free matter. One sees that the critical temperature
$T_c$ decreases slightly with increasing value of $L$ in the
neutrino-trapped matter, but it first decreases and then increases
with increasing $L$ in the neutrino-free matter. This complicated
behavior is due to the isospin and temperature effects on the
spinodal region and the relative neutron-proton abundance as shown
in Fig.~\ref{rhonrhop}. Our results thus indicate that for
neutrino-trapped neutron stars of temperatures higher than $12$ MeV
or for neutrino-free neutron stars of temperatures higher than $1.5$
MeV, there exists no inner crust if the value of $L$ is $70 \sim 80$
MeV. As newly-born hot neutron stars cool, the temperature at which
the inner crust can form thus depends on the density dependence of
the symmetry energy at subsaturation densities. Again, the magnitude
of the critical temperature for both neutrino-trapped and
neutrino-free matter is similar to those from Skyrme interactions
and relativistic mean field models~\cite{Duc08,Par09}.

The above results were obtained using the dynamical approach that
includes the effects of both the density gradient terms and the
Coulomb interaction. Neglecting these effects, the resulting
thermodynamical approach given by Eq.~(\ref{ther}) gives higher
values for both the transition density and pressure. This is
especially the case for the neutrino-trapped matter and/or for
smaller values of $L$ as more electrons are present in such hot
neutron stars and the effect due to the Coulomb interaction becomes
more important. Also, there are recently some studies on the
transition density in neutron stars using various nucleon-nucleon
interactions~\cite{Sur09,Vid09}. To see the effect of momentum
dependence in the nucleon-nucleon interaction on the transition
density and pressure, we have also repeated above calculations using
the momentum-independent MID interaction~\cite{Xu08}, which gives
the same equation of state for asymmetric nuclear matter but
different single-particle mean-field potential in  comparison with
the momentum-dependent MDI interaction used in the present study,
and we find that the momentum-dependent effect on the transition
density and pressure in hot neutron stars is small.

\section{Summary}
\label{summary}

We have studied the transition density and pressure at the boundary
that separates the liquid core from the inner crust of neutron stars
using the momentum-dependent MDI interaction in both the
neutrino-trapped matter and the neutrino-free matter at finite
temperatures, which are expected to exist during the early evolution
of neutron stars. In particular, we have investigated the effect of
nuclear symmetry energy by varying the parameter $x$ in the MDI
interaction from $1$ to $-1$, corresponding to the values $15<L<106$
MeV for the slope of nuclear symmetry energy at normal density that
were constrained by both the isospin diffusion
data~\cite{Tsa04,Che05a,LiBA05} and other experimental
observables~\cite{She10}.  We have found that the transition density
and pressure are larger in the neutrino-trapped matter than in the
neutrino-free matter. Furthermore, the transition density and
pressure are found to roughly decrease with increasing temperature
and $L$ for both the neutrino-trapped and the neutrino-free matter,
except that the transition pressure shows a complicated relation to
the temperature for the neutrino-trapped matter. Also, negative
values of the pressure at the transition density have been obtained,
which can be used to rule out a very stiff symmetry energy at
subsaturation densities. We have also studied the critical
temperature above which the inner crust (nuclear 'pasta' phase)
cannot be formed in newly-born neutron stars and found that it
depends sensitively on the density dependence of the nuclear
symmetry energy at subsaturation densities.

\begin{acknowledgments}

This work was supported in part by U.S. National Science Foundation
under Grant No. PHY-0758115, PHY-0652548 and PHY-0757839, the Welch
Foundation under Grant No. A-1358, the Research Corporation under
Award No. 7123, the Texas Coordinating Board of Higher Education
Award No. 003565-0004-2007, the National Natural Science Foundation
of China under Grant Nos. 10675082 and 10975097, MOE of China under
project NCET-05-0392, Shanghai Rising-Star Program under Grant No.
06QA14024, the SRF for ROCS, SEM of China, the National Basic
Research Program of China (973 Program) under Contract No.
2007CB815004.

\end{acknowledgments}


\begin{thebibliography}{99}

\bibitem{Cha08} N. Chamel and P. Haensel, Living
Rev. Relativity, \textbf{11}, 10 (2008).

\bibitem{Rus06} S. B. Ruster, M. Hempel, and J. Schaffner-Bielich, Phys. Rev. C \textbf{73}, 035804 (2006).

\bibitem{Lat00} J. M. Lattimer and M. Prakash, Phys. Rep. \textbf{333},
121 (2000).

\bibitem{Lat07} J. M. Lattimer and M. Prakash, Phys. Rep. \textbf{442}, 109
(2007).

\bibitem{LCK08} B. A. Li, L. W. Chen, and C. M. Ko, Phys. Rep. \textbf{464},
113 (2008).

\bibitem{She10} D. V. Shetty and S. J. Yennello, arXiv:1002.0313 [nucl-ex]

\bibitem{Tsa04} M. B. Tsang et al., Phys. Rev. Lett. \textbf{92}, 062701
(2004).

\bibitem{Che05a} L. W. Chen, C. M. Ko, and B. A. Li, Phys. Rev. Lett. \textbf{94%
}, 032701 (2005) ; Phys. Rev. C \textbf{72}, 064309 (2005).

\bibitem{LiBA05} B. A. Li and L. W. Chen, Phys. Rev. C \textbf{72}, 064611
(2005).

\bibitem{Tsa09} M. B. Tsang et al., Phys. Rev. Lett.
\textbf{102}, 122701 (2009).

\bibitem{Das03} C. B. Das, S. Das Gupta, C. Gale, and B. A. Li, Phys. Rev. C
\textbf{67}, 034611 (2003).

\bibitem{Kub07} S. Kubis, Phys. Rev. C \textbf{76}, 025801 (2007); Phys.
Rev. C \textbf{70}, 065804 (2004).

\bibitem{BPS71} G. Baym, C. Pethick and P. Sutherland, Astrophys. J. \textbf{170}, 299 (1971).

\bibitem{BBP71} G. Baym, H. A. Bethe and C. J. Pethick, Nucl. Phys. \textbf{A175}, 225 (1971).

\bibitem{Pet95a} C. J. Pethick and D. G. Ravenhall, Ann. Rev. Nucl. Part.
Sci. \textbf{45}, 429 (1995).

\bibitem{Pet95b} C. J. Pethick, D. G. Ravenhall and C. P. Lorenz, Nucl. Phys. \textbf{A584}, 675 (1995).

\bibitem{Oya07} K. Oyamatsu and K. Iida, Phys. Rev. C \textbf{75}, 015801 (2007).

\bibitem{Lin99} B. Link, R. I. Epstein, and J. M. Lattimer, Phys. Rev. Lett. \textbf{83},
3362 (1999).

\bibitem{XCLM09} J. Xu, L. W. Chen, B. A. Li, and H. R. Ma, Phys. Rev.
C \textbf{79}, 035802 (2009); Astrophys. J. \textbf{697}, 1549
(2009).

\bibitem{Bur86} A. Burrows and J. M. Lattimer, Astrophys. J.
\textbf{307}, 178 (1986).

\bibitem{Bet90} H. A. Bethe, Rev. Mod. Phys. \textbf{62}, 801 (1990).

\bibitem{Bur03} R. Buras, et al., Phys. Rev. Lett. \textbf{90},
241101 (2003).

\bibitem{BBLA79} H. A. Bethe, G. E. Brown, J. Applegate and J. M.
Lattimer, Nucl. Phys. \textbf{A324}, 487 (1979).

\bibitem{Bur88} A. Burrows,  Astrophys. J. \textbf{334}, 891 (1988).

\bibitem{Hor04} C. J. Horowitz, M. A.
P$\acute{e}$rez-Garc$\acute{i}$a and J. Piekarewicz, Phys. Rev. C
\textbf{69}, 045804 (2004).

\bibitem{Xu07} J. Xu, L. W. Chen, B. A. Li and H. R. Ma, Phys. Rev. C \textbf{75%
}, 014607 (2007).

\bibitem{Bog79} J. Boguta, Phys. Lett. \textbf{B106}, 255 (1979).

\bibitem{Lat91} J. Lattimer, et al, Phys. Rev. Lett. \textbf{66}, 2701
(1991).

\bibitem{Mar03} J. Margueron and P. Chomaz, Phys. Rev. C \textbf{67},
041602(R) (2003).

\bibitem{LiBA01} B. A. Li, et al, Phys. Rev. C \textbf{64}, 051303(R)
(2001);  Nucl. Phys. \textbf{A699}, 493 (2002).

\bibitem{LiBA97b} B. A. Li and C. M. Ko, Nucl. Phys. \textbf{A618}, 498 (1997).

\bibitem{Bar98} V. Baran, et al., Nucl. Phys. \textbf{%
A632}, 287 (1998).

\bibitem{Duc07} C. Ducoin, Ph. Chomaz and F. Gulminelli, Nucl. Phys. \textbf{%
A789}, 403 (2007).

\bibitem{Duc08} C. Ducoin, et al., Phys. Rev. C \textbf{78}, 055801 (2008).

\bibitem{Ava04} S. S. Avancini, et al, Phys. Rev. C \textbf{70},
015203 (2004).

\bibitem{Ava09} S. S. Avancini, et al., Phys. Rev. C \textbf{79},
035804 (2009).

\bibitem{Par09} H. Pais, A. Santos and C. Provid$\hat{e}$ncia, Phys. Rev.
C \textbf{80}, 045808 (2009).

\bibitem{Sur09} A. Sulaksono and Kasmudin, Phys. Rev. C \textbf{80},
054317 (2009).

\bibitem{Vid09} I. Vida$\tilde{n}$a, et al.,
Phys. Rev. C \textbf{80}, 045806 (2009).

\bibitem{Xu08} J. Xu, L. W. Chen, B. A. Li and H. R. Ma, Phys. Lett. \textbf{B650},
348 (2007); Phys. Rev. C \textbf{77}, 014302 (2008).

\end{thebibliography}
\end{document}